\documentclass[%
 preprint,
 amsmath,amssymb,
 aps,
]{revtex4-2}

\usepackage{graphicx}
\usepackage{dcolumn}
\usepackage{bm}

\begin{document}


\title{Nonreciprocal light propagation induced by a subwavelength spinning cylinder}

\author{Zheng Yang} \affiliation{Department of Physics, City University of Hong Kong, Tat Chee Avenue, Kowloon, Hong Kong, China}

\author{Yuqiong Cheng} \affiliation{Department of Physics, City University of Hong Kong, Tat Chee Avenue, Kowloon, Hong Kong, China}

\author{Neng Wang} \affiliation{College of Electronics and Information Engineering, Shenzhen University, Shenzhen 518052, Guangdong, China}

\author{Yuntian Chen} \affiliation{School of Optical and Electronic Information, Huazhong University of Science and Technology, Wuhan 430074, Hubei, China}

\author{Shubo Wang}\email {shubwang@cityu.edu.hk} \affiliation{Department of Physics, City University of Hong Kong, Tat Chee Avenue, Kowloon, Hong Kong, China} \affiliation{City University of Hong Kong Shenzhen Research Institute, Shenzhen 518057, Guangdong, China}

\date{\today}

\begin{abstract}
Nonreciprocal optical devices have broad applications in light manipulations for communications and sensing. Non-magnetic mechanisms of optical nonreciprocity are highly desired for high-frequency on-chip applications. Here, we investigate the nonreciprocal properties of light propagation in a dielectric waveguide induced by a subwavelength spinning cylinder. We find that the chiral modes of the cylinder can give rise to unidirectional coupling with the waveguide via the transverse spin-orbit interaction, leading to different transmissions for guided wave propagating in opposite directions and thus optical isolation. We reveal the dependence of the nonreciprocal properties on various system parameters including mode order, spinning speed, and coupling distance. The results show that higher-order chiral modes and larger spinning speed generally give rise to stronger nonreciprocity, and there exists an optimal cylinder-waveguide coupling distance where the optical isolation reaches the maximum. Our work contributes to the understanding of nonreciprocity in subwavelength moving structures and can find applications in integrated photonic circuits, topological photonics, and novel metasurfaces.
\end{abstract}

\maketitle


\section{Introduction}
Lorentz-reciprocal devices possess symmetric response with respect to the interchange of source and observation point, and the scattering matrix satisfies the symmetric condition $\mathbf{S}^{\mathrm{T}}=\mathbf{S}$ \cite{potton2004reciprocity, jalas2013and, sounas2017non}. Nonreciprocal devices breaking this symmetry, such as optical isolators\cite{auracher1975new, yokoi2000demonstration, huang2018nonreciprocal, maayani2018flying, shi2021optical} and  optical circulators\cite{kamal2011noiseless, scheucher2016quantum}, play critical roles in  full-duplex optical communications and invisible sensing. Various approaches have been proposed to achieve nonreciprocity, and they can be broadly divided into three categories based on their physical mechanisms: nonlinearity\cite{peng2014parity,khanikaev2015nonlinear, shi2015limitations,  lawrence2018nonreciprocal,lawrence2019nanoscale}, spatiotemporal modulations\cite{ yu2009complete,estep2014magnetic,Galiffi2019,guo2019nonreciprocal,Galiffi2022}, and external biasing\cite{khanikaev2010one, hadad2010magnetized, bi2011chip}. Nonlinear effects usually require high-intensity signals. Spatiotemporal modulations face challenges in achieving fast and robust effects at optical frequencies. External magnetic field biasing relies on the use of magnetic materials, which have significant loss at high frequencies and are incompatible with on-chip optical devices\cite{sounas2017non}. In contrast, synthetic external bias induced by structural rotation can give rise to robust and magnet-free nonreciprocal functionalities\cite{ sarma2015rotating,lannebere2016wave, lu2017optomechanically}. Conventional research about rotating optical structures mainly focuses on bulky cavities that support whispering gallery modes \cite{huang2018nonreciprocal, maayani2018flying, zhang2020breaking, jing2021nonreciprocal}. The physics cannot be directly extended to subwavelength nanophotonic systems where the spatial and polarization properties of light fields are inseparable. In particular, the spin and orbital degrees of freedom become strongly coupled with each other in subwavelength structures\cite{bliokh2015spin, wang2019arbitrary, chen2020chiral}. Therefore, it is of critical importance to explore the nonreciprocal properties and physics of subwavelength rotating structures.

In this paper, we numerically investigate the nonreciprocal properties of light propagation induced by a subwavelength spinning cylinder sitting above a dielectric waveguide. We show that the rotation breaks time-reversal symmetry and induces the Sagnac frequency splitting of two opposite chiral modes carrying opposite spin angular momentum. Under the excitation of the input guided wave, the fields of the chiral modes mainly couples to the guided wave propagating in one direction due to the transverse spin-orbit interaction of light\cite{bliokh2015spin}. This unidirectional coupling gives rise to different transmissions for the light input at opposite ends of the waveguide, the contrast of which corresponds to the optical isolation ratio. We conducted a detailed study of the dependence of the isolation ratio on various system parameters, including the order of the chiral modes, the spinning speed of the cylinder, and the cylinder-waveguide coupling distance. The results show that in general the isolation ratio can be increased by increasing the mode order and the spinning speed, and the maximum isolation ratio is obtained at an optimal separation distance between the cylinder and the waveguide. 

The paper is organized into four sections. In Section 2, we provide the formulations to describe the chiral mode properties of the spinning cylinder, and we discuss the frequency splitting of low-order chiral modes. In Section 3, we present the full-wave numerical results and discuss the nonreciprocal properties of the system. The nonreciprocal phenomena are understood based on the transverse spin-orbit interaction of the guided wave. In addition, we discuss the effects of various system parameters on the isolation ratio. We then draw the conclusion in Section 4.

\section{Formulations of the spinning cylinder}
The electromagnetic properties of a spinning cylinder can be characterized by the Minkowski constitutive relations\cite{minkowski1908grundgleichungen}:\begin{equation}
\begin{aligned}
&\mathbf{D}+\mathbf{v} \times \frac{\mathbf{H}}{c^{2}}=\boldsymbol{\varepsilon}\cdot(\mathbf{E}+\mathbf{v} \times \mathbf{B}), \\
&\mathbf{B}+\mathbf{E} \times \frac{\mathbf{v}}{c^{2}}=\boldsymbol{\mu} \cdot(\mathbf{H}+\mathbf{D} \times \mathbf{v}),
\end{aligned}
\end{equation}
where $\mathbf{E}, \mathbf{D}, \mathbf{B}, \mathbf{H}$ are the electromagnetic fields, $c$ is the speed of light in vacuum, $\mathbf{v}$ is the linear velocity that depends on the radial distance, $\boldsymbol{\varepsilon}$  and  $\boldsymbol{\mu}$ are the permittivity and permeability tensors of the stationary cylinder, respectively. Equation (1) indicates that the electric and magnetic fields will couple with each other due to the spinning motion of the cylinder, and the spinning cylinder behaves like a bi-anisotropic medium. The above constitutive relations can be re-written as 
\begin{equation}
\left[\begin{array}{l}
\mathbf{D} \\
\mathbf{B}
\end{array}\right]=\left[\begin{array}{cc}
\boldsymbol{\varepsilon}^{\prime} & \boldsymbol{\chi}_{\mathrm{em}} \\
\boldsymbol{\chi}_{\mathrm{me}} & \boldsymbol{\mu}^{\prime}
\end{array}\right]\left[\begin{array}{l}
\mathbf{E} \\
\mathbf{H}
\end{array}\right],
\end{equation}
where $ \boldsymbol{\chi}_{\mathrm{me}}$ and $ \boldsymbol{\chi}_{\mathrm{em}}$ characterize the bi-anisotropic properties of the medium. The elements in the above constitutive tensor satisfy $\boldsymbol{\varepsilon}^{\prime}=\boldsymbol{\varepsilon}^{\prime \mathrm{T}}, \boldsymbol{\mu}^{\prime}=\boldsymbol{\mu}^{\prime \mathrm{T}}$ and $\boldsymbol{\chi}_{\mathrm{em}}=\left(\boldsymbol{\chi}_{\mathrm{me}}{ }^{*}\right)^{\mathrm{T}}=\boldsymbol{\chi}_{\mathrm{me}}^{\mathrm{T}}$, which indicates that the spinning motion breaks time-reversal symmetry and turns the cylinder into a nonreciprocal medium\cite{shi2019gauge, shi2021optical, shi2021robust}. Although the above formulations apply to general cylinders with cylindrical symmetry, in what follows we will focus on a dielectric cylinder that is isotropic and homogeneous when it is stationary.

The nonreciprocal properties can be significantly amplified by the resonance modes of the cylinder. Under the TE polarization, the eigenmodes of the cylinder can be expressed as $E_{z}=E_{z}(r) \exp (i m \theta)$, where $m$ is an integer denoting the azimuthal quantum number of the modes and $\theta$ is the azimuthal angle in the cylindrical coordinate system. The fields of the modes with $m>0$ and $m<0$ rotate in clockwise (CW) and counter-clockwise (CCW) directions with respect to the $z$ axis, respectively. Thus, they are chiral modes carrying orbital angular momentum in $z$ direction. For a stationary cylinder, the CW and CCW modes with $\pm m$ are degenerate due to the mirror symmetry and time-reversal symmetry. This degeneracy disappears when the cylinder rotates with non-zero speed as a result of the Sagnac effect \cite{POST1967}.  It also can be understood as a result of the synthetic gauge field induced by the rotation of the cylinder\cite{shi2019gauge}. At a low rotation speed $\Omega R \ll c$, where $\Omega$ is the angular velocity and $R$ is the radius of the cylinder, the Sagnac frequency splitting between a pair of chiral modes with $\pm m$ is\cite{shi2019gauge}: \begin{equation}
\Delta \omega=\frac{2 m\Omega\left(\varepsilon_{\mathrm{r}} \mu_{\mathrm{r}}-1\right) }{\varepsilon_{\mathrm{r}} \mu_{\mathrm{r}}},
\end{equation}
where $\varepsilon_{r}$ and $\mu_{r}$ are the relative permittivity and relative permeability of the stationary cylinder, respectively. Equation (3) indicates that the frequency splitting is proportional to the azimuthal quantum number $m$ and the angular velocity $\Omega$. 

\section{Results and discussion}
\subsection{Scattering properties of the isolated spinning cylinder}

\begin{figure}[b]
\centering\includegraphics[width=\linewidth]{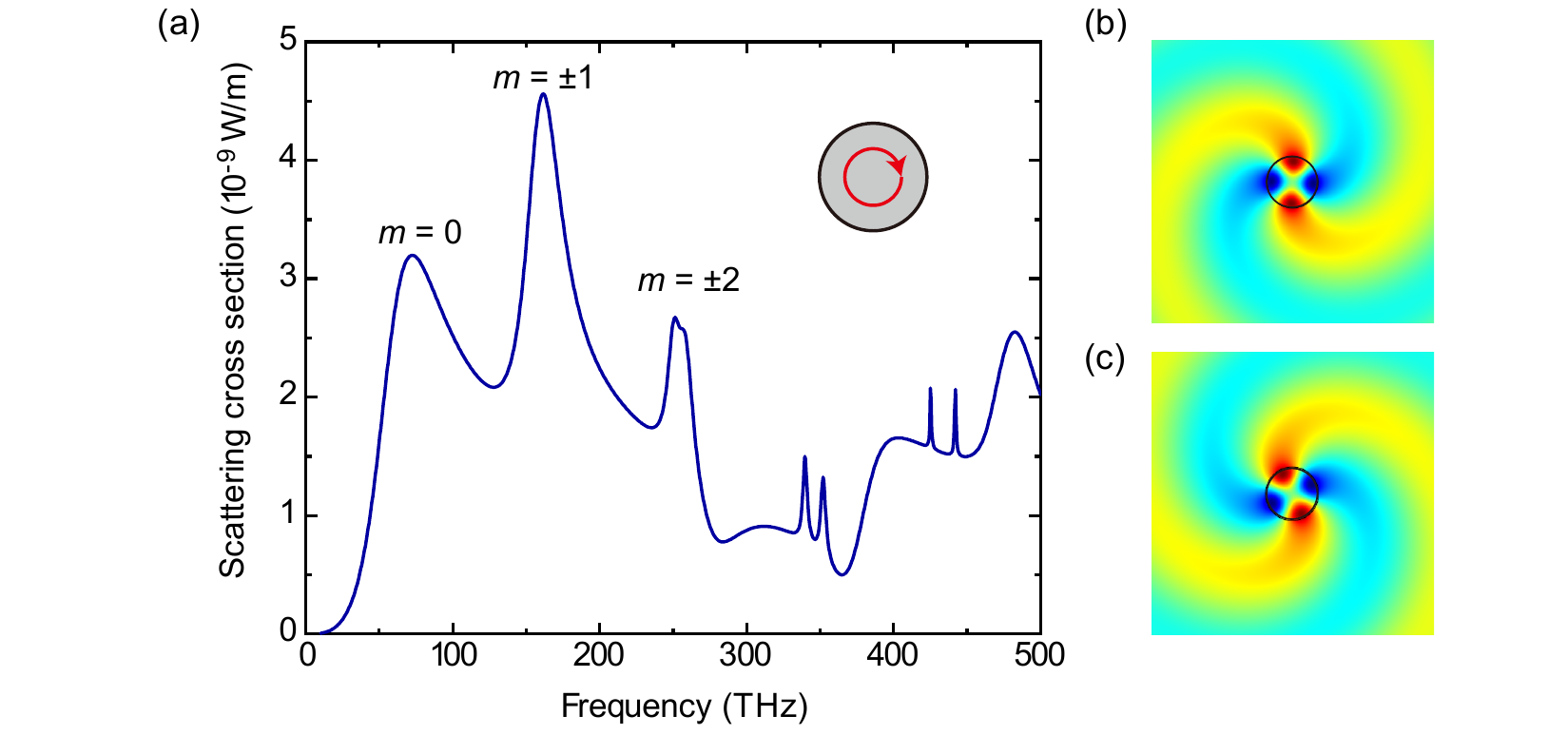}
\caption{Scattering cross section of the isolated spinning cylinder with spinning speed $\Omega R / c=0.01$. (b),(c) Eigen electric field of the chiral mode with $m=+2$ and $m=-2$, respectively.}
\label{fig1}
\end{figure}

We first explore the scattering properties of the isolated spinning cylinder made of silicon with $\varepsilon_r=11.9$ and $\mu_r=1$ (shown in the inset of Fig. 1(a)). We implement the constitutive relations in Eq. (2) by using a finite-element package COMSOL Multiphysics\cite{RN16}, and then we calculate the scattering cross section of the cylinder with the normalized spinning speed $\Omega R / c=0.01$. Figure 1 shows the scattering cross section of the spinning cylinder under the excitation of a TE-polarized plane wave with electric field amplitude 1 V/m. The peaks in the scattering spectrum correspond to the chiral modes of the cylinder and have been labelled by the corresponding $m$ value. The $m=0$ mode involves no degeneracy and thus no frequency splitting. The frequency splitting of $m=\pm 1$ modes is small and does not manifest in the scattering spectrum due to the relatively low quality factors of the modes. For the chiral modes with $|m| \geq 2$, two peaks appear in the scattering spectrum due to the Sagnac frequency splitting. Each pair of the peaks corresponds to a pair of the CW and CCW modes. It is obvious that the Sagnac frequency splitting is larger for higher-order chiral modes with a larger azimuthal quantum number $m$, agreeing with Eq. (3). Figure 1(b) and (c) show the eigen electric field of the CW ($m=+2$) and CCW ($m=-2$) chiral modes, respectively, where the chiral feature of the modes and their orbital angular momentum are clearly observed. 

\subsection{Nonreciprocal properties of guided waves induced by the spinning cylinder}
\begin{figure}[tb]
\centering\includegraphics[width=\linewidth]{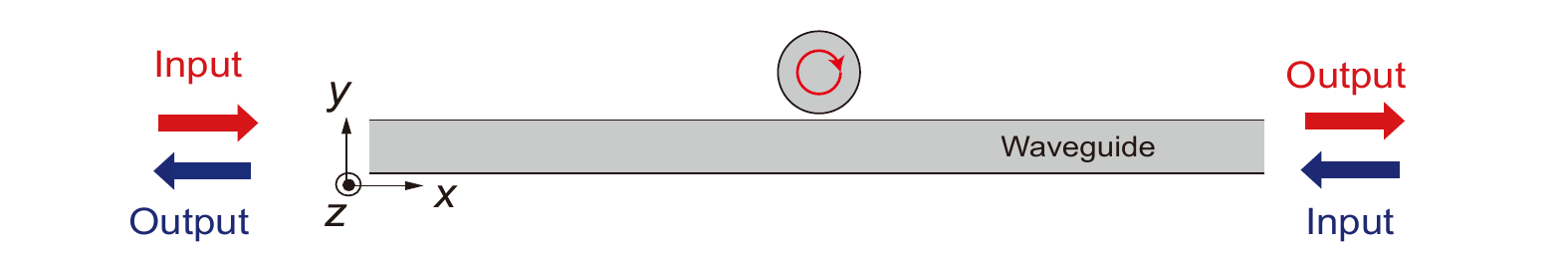}
\caption{Schematic of the nonreciprocal system composed of a spinning cylinder sitting on a slab waveguide. The cylinder has a radius of $R=200 \mathrm{~nm}$ and spins with angular velocity $\Omega$. The thickness of the waveguide is $260 \mathrm{~nm}$. The gap distance between the waveguide and the cylinder is $d$. }
\label{fig2}
\end{figure}

\begin{figure}[tb]
\centering\includegraphics[width=0.9\linewidth]{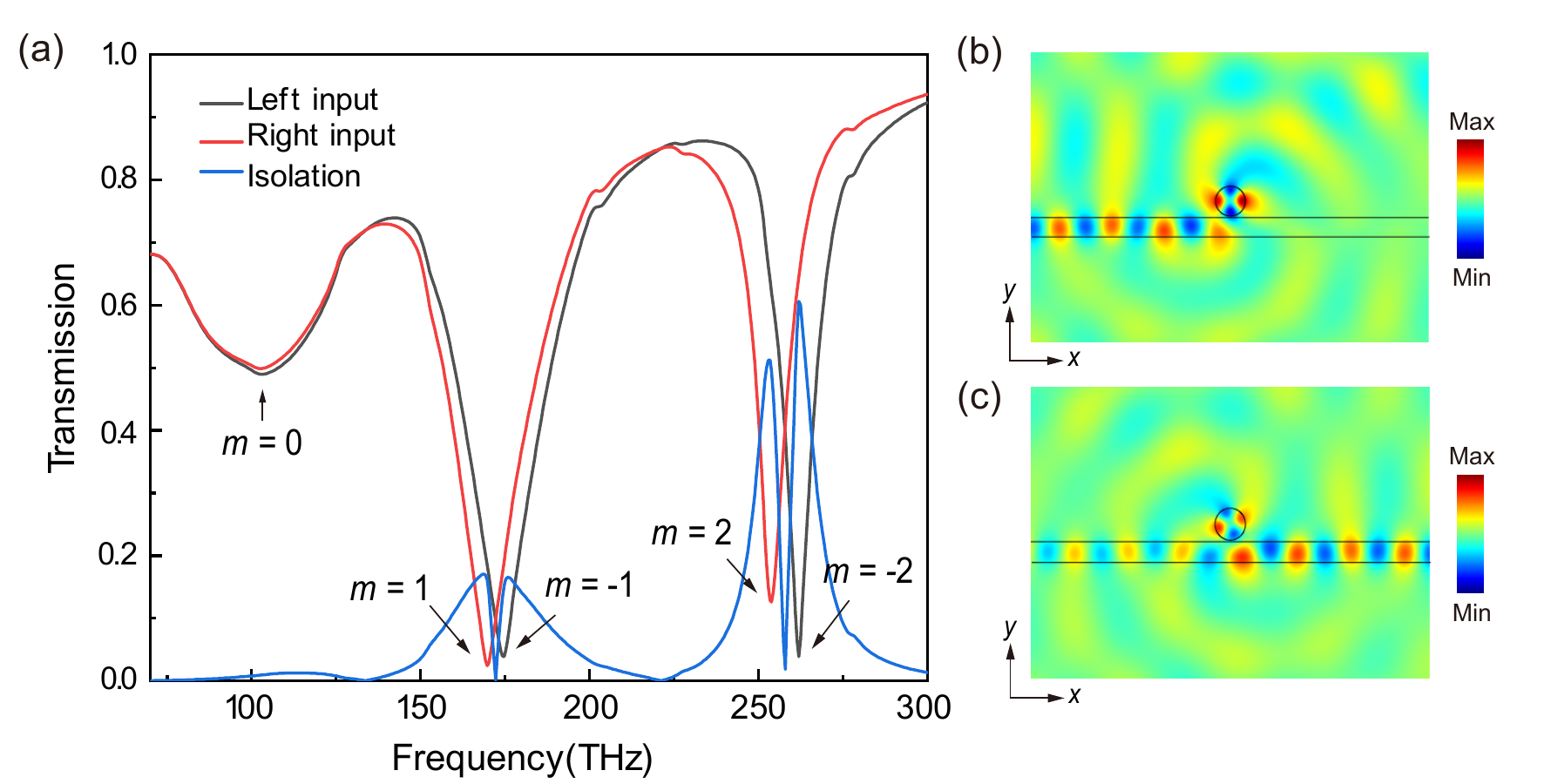}
\caption{(a) Transmission and isolation ratio of the fundamental TE waveguide mode at the frequencies of the low-order chiral modes. The blue line corresponds to the absolute difference of the black and red lines. (b), (c) $E_{z}$ field under port excitation from the left and right input, respectively. We set $d=20 \mathrm{~nm}$ and $\Omega R/c=0.01$. }
\label{fig3}
\end{figure}

The spinning cylinder can be employed to realize nonreciprocal propagation of light. We consider the two-dimensional (2D) system shown in Fig. 2, where the spinning cylinder ($\varepsilon_{\text{r}}=11.9$) sits above a slab waveguide made of silicon dioxide ($\varepsilon_{\text{wg}}=4$). The gap distance between the cylinder and the waveguide is $d$. The cylinder has a radius $R=200 \mathrm{~nm}$ and spins in the clockwise direction with angular velocity $\mathbf{\Omega}=\Omega \hat{\mathrm{z}}$. The waveguide has a thickness of $260 \mathrm{~nm}$, and only its fundamental TE mode can be excited at the considered frequencies. 

Under the excitation of the guided wave, the chiral modes of the cylinder will couple with the waveguide modes, which will affect the transmission of the guided wave. Figure 3(a) shows the transmission spectra of the fundamental TE guided waves input from the left (black) and right (red) ends of the waveguide, where we set $d=20 \mathrm{~nm}$ and $\Omega R/c=0.01$. We notice that the transmissions are different in a wide range of frequencies due to the nonreciprocal properties of the spinning cylinder. The transmissions manifest three dips in the considered frequency range, which are attributed to the $ m=0, \pm 1, \pm 2$ chiral modes. The blue line denotes the transmission contrast, i.e., optical isolation ratio, defined as the absolute difference between the black and red lines. As seen, the isolation ratio is significantly enhanced at the resonance frequencies of the chiral modes. The isolation ratio is larger at the chiral modes with a larger  $m$ value, and a maximum about $60 \%$ appears at 261.9 THz. This is because that the chiral modes with larger $m$ have larger quality factors and larger frequency splitting according to Eq. (3). Consequently, the overlap in the transmission spectra of CW and CCW modes is smaller, leading to larger transmission contrast and larger isolation ratio. Figure 3(b) and (c) show the electric field at the 261.9 THz for the left input and right input, respectively. As seen, the light input from the left end excites the CCW chiral mode and is almost completely blocked. In contrast, the light input from the right end excites the CW chiral mode and partially passes through the waveguide. This demonstrates the nonreciprocal propagation of light induced by the spinning motion of the cylinder.

\begin{figure}[tb]
\centering\includegraphics[width=\linewidth]{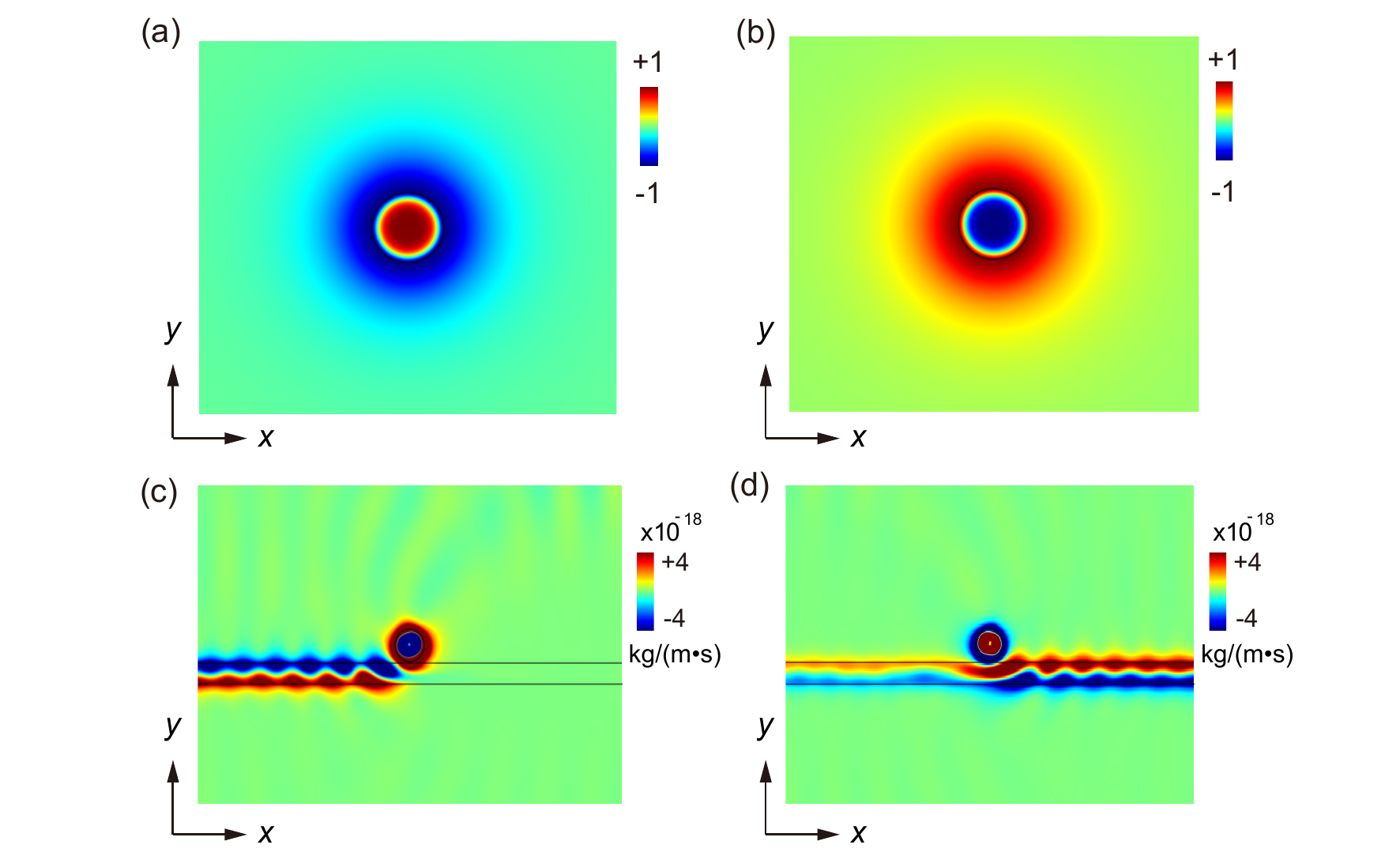}
\caption{(a), (b) Stokes parameter $S_{3}$ of the magnetic field for a pair of chiral modes with $m=+2$ and $m=-2$, respectively. (c), (d) Transverse spin density in the system under left excitation and right excitation, respectively.}
\label{fig4}
\end{figure}

The nonreciprocal properties of the system can be understood based on the transverse spin-orbit interaction of light\cite{petersen2014chiral, bliokh2015spin, bliokh2015quantum, wang2019arbitrary, shi2021optical}. In addition to orbital angular momentum, the chiral modes of the cylinder also carry spin angular momentum, which can be characterized by the Stokes parameter  \( S_{3} \) of the magnetic field. Figure 4(a) and 4(b) show \( S_{3} \) for the chiral modes of the isolated spinning cylinder with  $m=+2$ and $m=-2$,  respectively. As seen,  \( S_{3} \) is closely related to the circulating direction of the chiral mode, and it has opposite signs for the fields inside and outside the cylinder. For the mode of $m=+2$ in Fig. 4(a), the Stokes parameter inside the cylinder approaches $S_{3}=+1$,  while it takes the value of $S_{3}=-1$ outside the cylinder and near the boundary. The sign of the Stokes parameter is reversed for the mode of  $m=-2$,  as shown in Fig. 4(b). When the cylinder is near to the waveguide, the input guided wave will excite the chiral modes, and the evanescent wave of the chiral modes will couple to the waveguide. This excitation and coupling depend on the spin of the chiral modes and the spin of the guided wave. Figure 4(c) and 4(d) show the magnetic spin density $\frac{\mu_0}{4\omega}\text{Im}\left[\mathbf{H}^{*} \times \mathbf{H}\right]$ for light input from the left and right sides of the waveguide, respectively. We see that the guided waves propagating in $+x$ and $-x$ directions carry opposite spin in $z$ direction, and the spin near the upper and lower surfaces of the waveguide has opposite sign\cite{bliokh2015quantum}. When the light is input from left with $k_{x}>0$ (right with  $k_{x}<0$), it will excite the CCW (CW) chiral mode with negative (positive) spin inside the cylinder. The evanescent field of the chiral mode will couple back to the waveguide, and this coupling is unidirectional due to the spin-momentum locking property of the guided wave\cite{wang2019arbitrary,shi2021optical}. In the case of Fig. 4(c) (Fig. 4(d)), the chiral mode couples to the guided wave propagating to right (left) side, which interferes with the input light and suppresses the transmission\cite{shi2021optical}. The suppression of transmission is weaker in Fig. 4(d) because the CW mode is off resonance at this frequency due to the Sagnac frequency splitting. 

\begin{figure}[tb]
\centering\includegraphics[width=\linewidth]{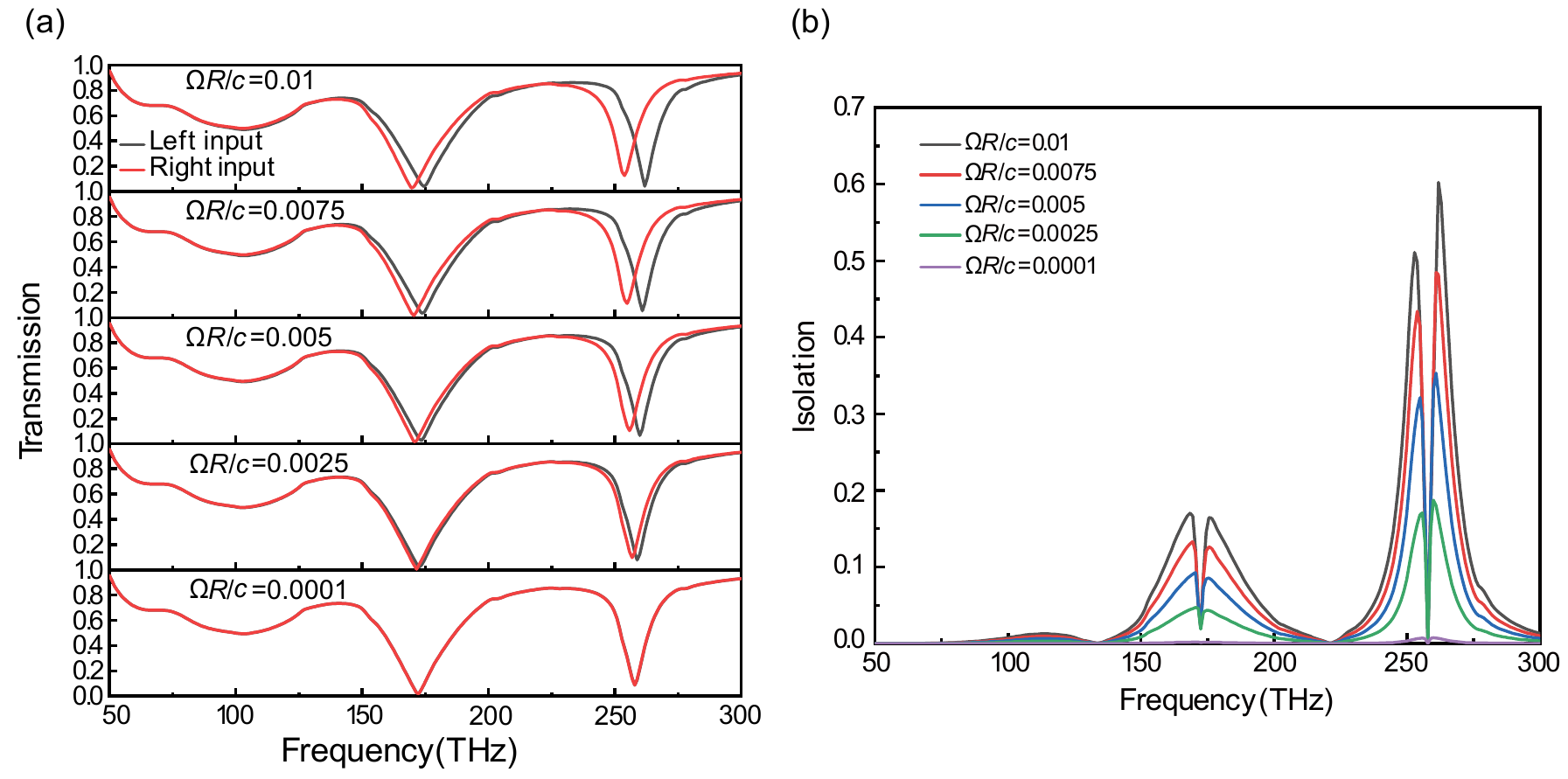}
\caption{(a) Transmission and (b) isolation ratio of the waveguide for different rotation speeds of the cylinder. The gap distance between the cylinder and the waveguide is fixed at $d=20$ nm.}
\label{fig5}
\end{figure}

To further explore the nonreciprocal properties of the system, we computed the transmission of the guided wave for various spinning speeds of the cylinder, fixing the gap distance $d=20 $ nm. The numerical results for rotation speeds $\Omega R/c=0.001,0.0025,0.005,0.0075,0.01$ are shown in Fig. 5(a). The black (red) line denote the transmission of guided wave input from the left (right) end of the waveguide. As seen, the spectral separation of the transmission dips corresponding to the CW and CCW chiral modes increases as the rotation speed is increased. This can be understood based on Eq. (3). The frequency splitting is proportional to the spinning speed. A higher speed leads to a larger frequency splitting and less overlap of the CW and CCW chiral modes in the frequency spectrum. Therefore, the isolation ratio  becomes larger when $\Omega R/c$ is larger. This is indeed what we observe in Fig. 5(b), which shows the isolation ratio for different rotation speeds of the cylinder,  corresponding to the cases in Fig. 5(a). We notice that the considered system can realize an isolation ratio about $60 \%$ at 261.9 THz (corresponding to the frequency of the CCW mode) with normalized spinning speed $\Omega R / c=0.01$.

Another important parameter that can affect the nonreciprocity of the system is the gap distance between the cylinder and the waveguide. We investigated the effect of different gap distances on the transmission and isolation ratio, fixing the spinning speed  $\Omega R / c=0.01$. Figure 6(a) shows the transmission of the guided wave input at the left (black) and right (red) end of the waveguide. We notice that, as the distance increases from $d=10$ nm to $d=150$ nm, the resonance induced transmission dips first 
decrease and then increase. Consequently, the isolation ratios at the resonance frequencies reach maximum values at an intermediate distance $d$. This is confirmed by the numerical results in Fig. 6(b), which shows the isolation ratio corresponding to the cases of Fig. 6(a). As seen, the isolation ratio first increases and then decreases when the distance increases from $d=10$ nm to $d=150$ nm. For the CW mode at the lower frequency, the maximum isolation ratio appears at the distance $d=50$ nm (corresponding to the blue line). For the CCW mode at the higher frequency, the maximum isolation ratio appears at the distance $d=20$ nm (corresponding to the green line). The existence of an optimal distance can be understood as follows. At a large distance, the coupling between cylinder and waveguide is weak due to the rapid decay of the evanescent waves, thus, the nonreciprocity induced by spinning motion can hardly affect the transmission of the guided wave. At a small distance, the presence of the waveguide breaks the cylindrical symmetry and induces strong coupling between the CW and CCW chiral modes, leading to a mix of opposite spins and thus excitation of guided waves propagating in both directions. This further reduces the unidirectionality and hence the isolation ratio. 

\begin{figure}[tb]
\centering\includegraphics[width=1\linewidth]{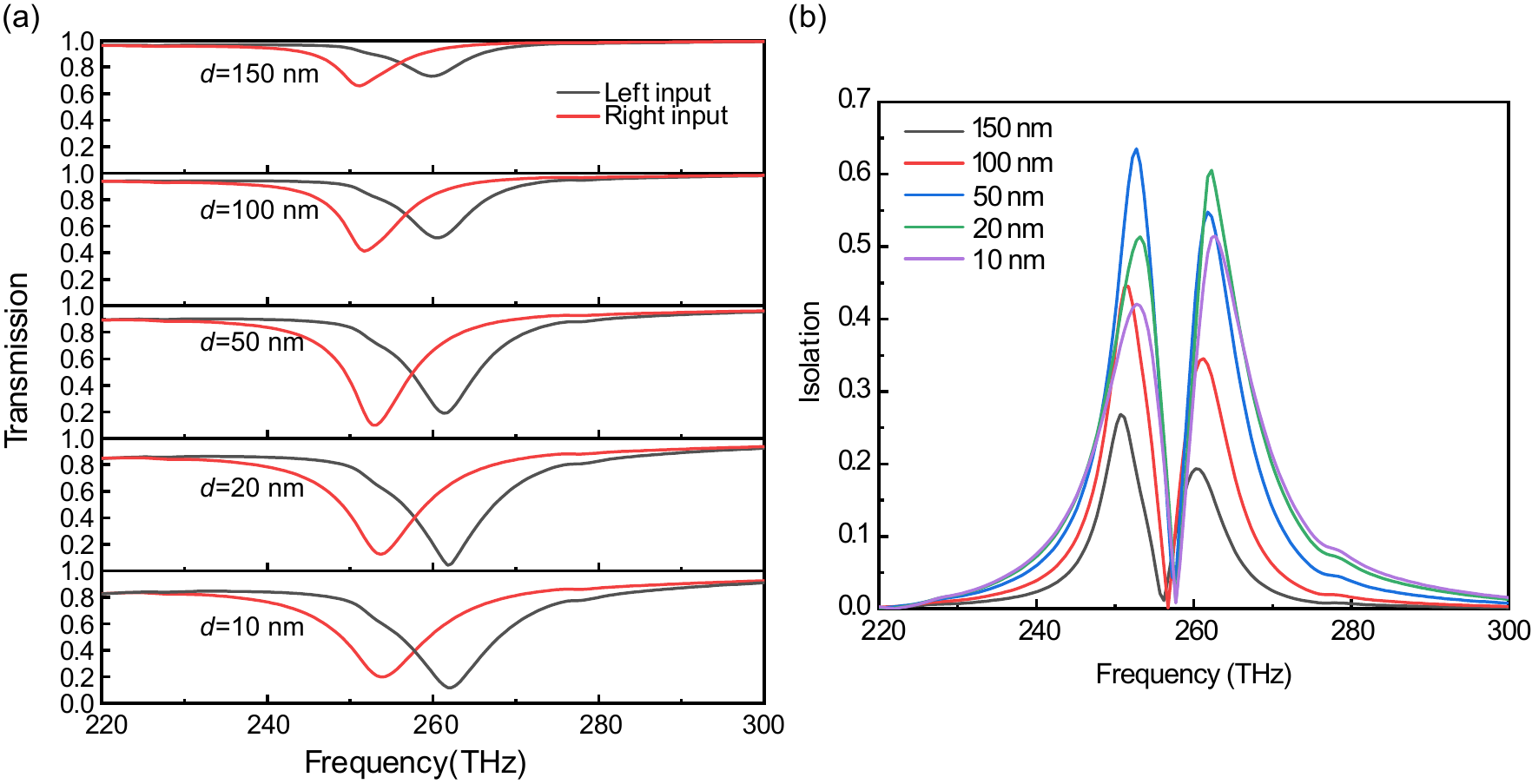}
\caption{(a) Transmission and (b) isolation ratio of the waveguide for different separation distances between the cylinder and the waveguide. The spinning speed of the cylinder is fixed at $\Omega R / c=0.01$.}
\label{fig6}
\end{figure}

\section{Conclusion}
In conclusion, we have investigated the nonreciprocal properties of light propagation in a dielectric waveguide induced by a subwavelength spinning cylinder. The spinning motion breaks the time-reversal symmetry and turns the cylinder into a bi-anisotropic medium. The frequency splitting induced by the spinning motion allows selective excitation of the CW/CCW chiral mode at one frequency, which can unidirectionally couple to the waveguide mode due to spin-momentum locking. This can enhance the nonreciprocity and give rise to high isolation ratio of the guided light. In addition, we studied the dependence of optical isolation on several important parameters, including the mode order, the spinning speed of cylinder, and cylinder-waveguide distance. We discover that the isolation ratio is generally higher for higher-order chiral modes and a larger spinning speed. Importantly, there is an optimum cylinder-waveguide distance where the isolation ratio reaches the maximum. Our study contributes to the understanding of nonreciprocal properties of moving optical structures. The results can find applications in designing on-chip nonreciprocal devices, novel metamaterials and metasurfaces.

\section*{Acknowledgements}
The work described in this paper was supported by the Research Grants Council of the Hong Kong Special Administration Region, China (CityU 11301820) and National Natural Science Foundation of China (11904306, 11904237, 12174263).

\bibliography{apssamp}

\end{document}